\documentclass{article}
\usepackage{graphicx}

\begin{document}
\def\cz{convection zone}
\def\mc{meridional circulation}

\title{The irregularities of the sunspot cycle and their theoretical modelling}
\author{Arnab Rai Choudhuri}

\maketitle

\begin{abstract}
The 11-year sunspot cycle has many irregularities, the most prominent amongst
them being the grand minima when sunspots may not be seen for several cycles.
After summarizing the relevant observational data about the irregularities,
we introduce the flux transport dynamo model, the currently most successful
theoretical model for explaining the 11-year sunspot cycle.  Then we analyze
the respective roles of nonlinearities and random fluctuations in creating the
irregularities.  We also discuss how it has recently been realized that the
fluctuations in meridional circulation also can be a source of irregularities.
We end by pointing out that fluctuations in the poloidal field generation and
fluctuations in \mc\ together can explain the occurrences of grand minima.
\end{abstract}

\section{Introduction}

The number of sunspots seen on the solar surface rises and falls with a period of
about 11 years. This 11-year cycle of sunspots is one of most intriguing natural
cycles which is affecting our lives in many ways as our society becomes more
dependent on technology. Violent explosions known as solar flares occur more
frequently when there are more sunspots.  Apart from producing the beautiful
polar aurorae, a large flare can disturb the ionosphere causing disruptions in
radio communication, can damage electronics in man-made satellites, can make
airlines flights near geomagnetic poles particularly hazardous and can even 
trip power grids.  On 13 March 1989, a large part of eastern Canada had a power
blackout caused by a powerful solar flare.

\begin{figure}
\center
\includegraphics[width=12cm]{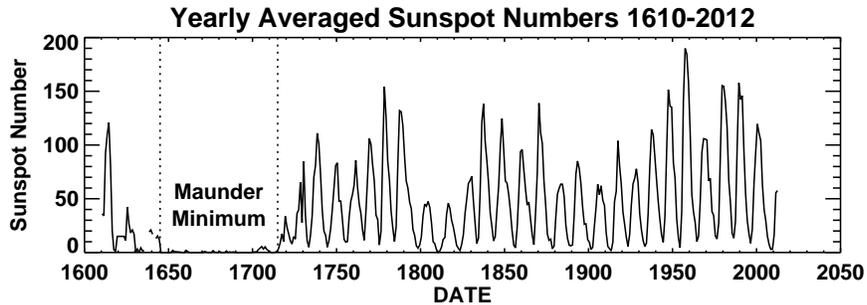}
  \caption{A plot of the yearly averaged sunspot number from 1610 to the present time.}
\end{figure}

Figure~1 shows the sunspot number as a function of time from 1610. Galileo and some of
his contemporaries were the first scientists to study sunspots
systematically.  The initial entries in Figure~1 are based on
their records.  Then, for nearly a century, sunspots were
rarely seen---a period known as the {\it Maunder minimum}.  Afterwards
the sunspot number has varied periodically with a rough period
of about 11 years, although we see a considerable amount of irregularity.
Some cycles are stronger than the average and some are weaker.

After the discovery of the sunspot cycle by Schwabe in 1843 [1], for a long
time there was no theoretical explanation for it. When Hale discovered in 1908 [2]
that a sunspot is a site of a concentrated magnetic field (about 0.3 T, only
a little bit weaker than the strongest magnetic fields produced in our laboratories
by large electromagnets), it became clear that the 11-year sunspot cycle
is essentially the magnetic cycle of the Sun.
It may be mentioned that Hale's discovery of magnetic fields in sunspots
was a truly momentous discovery in the history of physics because this
was the first time somebody found a conclusive evidence of large-scale
magnetic fields outside the Earth's environment.  Now we know that magnetic
fields are ubiquitous in the astronomical universe.

It is now generally accepted that a magnetohydrodynamic (MHD) process known as the dynamo
process is responsible for generating magnetic fields in astrophysical
systems.  The foundations of dynamo theory were laid down in a 1955 classic
paper by Parker [3], in which he derived the dynamo equation arising out of
MHD turbulence subject to rotation.  Afterwards, Steenbeck, Krause and R\"adler [4]
developed the mean field formalism of dynamo theory in a more systematic way.

The particular dynamo process responsible for producing the 11-year sunspot
cycle is called the flux transport dynamo process.  Invoking some early ideas due to
Babcock [5] and Leighton [6], the flux transport dynamo theory was first
formulated by Wang, Sheeley and Nash [7], Choudhuri, Sch\"ussler and Dikpati [8]
and Durney [9]. This theory has been remarkably successful in providing
theoretical explanation of various aspects of the sunspot cycle.   At first,
efforts were focussed on explaining regular aspects of the sunspot cycle.
After the successful modelling of the regular aspects, the thrust of research in
the last few years has been to apply the flux transport dynamo model to study
the irregularities of the sunspot cycle.  

An earlier review by the present author [10] summarized the basic
observational data about the sunspot cycle and then discussed how the flux transport
dynamo model was developed to explain these observational data.  Although we
shall briefly summarize the salient features of the flux transport dynamo, we
do not want to repeat the full discussions of the previous review. So we would
urge the readers to read this previous review before reading the present review.
The present review can be regarded as a continuation of the previous review. The
main aim of the present review will be to discuss how the irregularities of the
sunspot cycle are modelled with the flux transport dynamo. Although a little bit
of discussion of this subject can be found at the end of the previous review [10], some
very important developments took place in this field after that review
was written.  These very recent developments will be highlighted throughout the
present review.

\section{Some aspects of observational data}

The earlier review [10] provided a summary of the regular periodic
aspects of the sunspot cycle (Hale's polarity law, butterfly diagram).  So we not
discuss those topics here.  We merely focus our attention on the irregularities of the sunspot
cycle.  If all the irregularities were really `irregular' in the true sense, then
it would have been very difficult to develop any theoretical understanding about them.
However, one can discern certain patterns within the irregularities which give us
valuable clues how these irregularities may arise and how they can be modelled
theoretically. 

To discover patterns within the irregularities of the sunspot cycle, 
one would like to have as much data
about the irregularities as possible, so that statistical inferences
become meaningful.  We have actual sunspot records for about four centuries,
although the records become less reliable as we go earlier than the nineteenth
century. One important question is whether we have other proxies of sunspot
activity through which we can infer about sunspot cycles in the past even
without actual sunspot records.  When the sunspot activity is low, the magnetic
field in the solar wind becomes weaker, allowing more cosmic ray particles to
reach the Earth and to produce more of the radioactive nuclei $^{10}$Be and $^{14}$C
by interacting with air molecules. If we can infer what the concentrations of    
$^{10}$Be and $^{14}$C in the atmosphere were at earlier times, then from that
we can reconstruct a history of sunspot cycles in the past.  The atmospheric concentration
of $^{14}$C in the past can be inferred by analyzing old tree rings, whereas
the atmospheric concentration of $^{10}$Be in the past can be inferred from the
polar ice cores which have formed over many years.  It has now been possible
to reconstruct the history of sunspot activity for the past 11,000 years.

At the first sight, the strengths of different sunspot cycles as seen in Figure~1
may appear to vary randomly. Let us first discuss if there are any long-term
patterns.  Sunspot cycles have been numbered from the middle of the eighteenth
century, the present cycle being cycle~24. For several cycles from cycle~10,
the odd cycle has been stronger than the previous even cycle, a pattern at last
broken by cycle~23 which turned out to be weaker than cycle~22. This is called
the Gnevyshev--Ohl rule [11], though departures from this rule are known.  Apart from
this two-cycle pattern, it is often claimed that there a modulation of cycle
amplitudes involving eight cycles, often called the Gleissberg cycle.  From the
limited data we have, it is very difficult to either prove or disprove the
existence of the Gleissberg cycle.  What is clear, however, is that sometimes
the sunspot activity may almost disappear for many years and several cycles
may go missing, like what happened during the Maunder minimum.  Such events
are called grand minima.  On reconstructing the sunspot activity for several
millenia, it is now clear that the Maunder minimum was not unique.  It is
estimated that there have been 27 such grand minima during the last 11,000 years [12].

Apart from these patterns involving the amplitudes of different cycles, there are
other patterns within the irregularities of sunspot cycles.  The earlier review
[10] discussed in detail the possible correlation between the polar
fields during the sunspot minima and the strengths of next cycles.  If such a
correlation does exist (which seems to be the case from the limited data we have),
then that gives a powerful tool for predicting the strength of a sunspot cycle
before its beginning, once we know the strength of the polar field during the
previous sunspot minimum.  The last interesting pattern to which we wish to
draw the readers' attention is what is called the Waldmeier effect [13].  It appears
that strong cycles rise fast, whereas weak cycles rise more slowly.  In other
words, there is an anti-correlation between the rise times of the cycles and their
strengths.  

Within the last few years, attempts are being made to explain these patterns
of irregularities with the flux transport dynamo model.  After discussing the
basic model in the next section, we shall come to the theoretical modelling of
irregularities from \S~3.

\section{Flux transport solar dynamo}

We now give a very brief summary of the flux transport dynamo model of the sunspot
cycle.  We emphasize again that this discussion is not meant to be self-explanatory.
It is not meant to be accessible to readers without any previous knowledge of the
subject.  Readers without any previous knowledge are urged to read the previous review [10]
before proceeding further. 

The toroidal and the poloidal components of the Sun's magnetic field are supposed
to sustain each other through a feedback loop.  The differential rotation of the
Sun (which is now fully mapped by helioseismology) stretches out the poloidal field
to produce the toroidal field.  This primarily takes place at the bottom  of the solar
convection zone (at $r = 0.7 R_{\odot}$) where the differential rotation is concentrated.
To complete the dynamo loop, the poloidal field has to be generated back from this toroidal
field.  How this happens is more subtle.  The original idea of Parker [3] and
Steenbeck, Krause and R\"adler [4]---often called the $\alpha$-effect---was 
that the toroidal field is twisted by the
helical turbulence of the convection zone to produce the poloidal field.  This is
possible only if the toroidal field does not have energy density more than the energy
density of turbulence.  The condition for this is that the toroidal field should not
be stronger than $10^4$ G. The idea of the toroidal field being twisted by helical
turbulence had to be questioned when detailed calculations of the rise of the toroidal
field by magnetic buoyancy to form sunspots were out on the basis of the thin flux
tube equation [14--15]. The simulations of Choudhuri and Gilman [16],
Choudhuri [17], D'Silva and Choudhuri [18] and Fan, Fisher and DeLuca [19] suggested
that the toroidal field at the bottom of the convection zone has to be as strong as
$10^5$ G in order to match different aspects of observations. The $\alpha$-effect cannot
operate on such a strong field.

An alternative idea of the poloidal field generation goes back to Babcock [5]
and Leighton [6]. The toroidal field rising due to magnetic buoyancy produces
bipolar sunspots on the solar surface with tilts caused by the Coriolis force---an
effect known as Joy's law. When a tilted bipolar sunspot decays, the two opposite
magnetic polarities spread preferentially in slightly different latitudes. Many of
us now believe that the poloidal field generation in the solar dynamo takes place due
to this Babcock--Leighton mechanism.  The Sun has a meridional circulation which is
poleward near the surface and advects this poloidal field poleward [20--23]. This meridional
circulation also plays a crucial role in the solar dynamo.  The kind of dynamo in
which the poloidal field is generated by the Babcock--Leighton mechanism and the
meridional circulation plays a critical role is called the flux transport dynamo.

\begin{figure}
\center
\includegraphics[width=7cm]{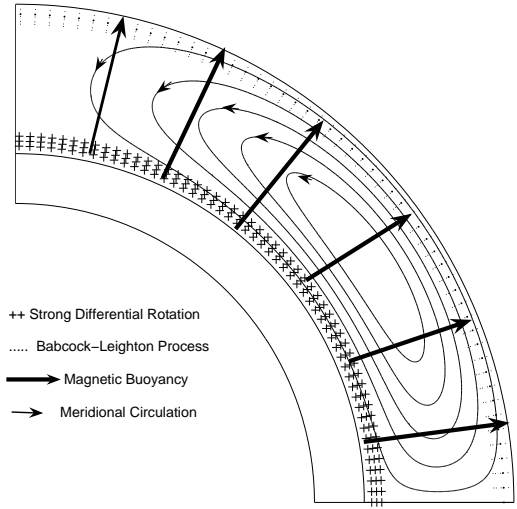}
  \caption{A cartoon explaining how the flux transport dynamo works.}
\end{figure}

Figure~2 is a cartoon explaining how the flux transport dynamo operates within the
solar convection zone. The toroidal field is generated at the bottom
of the convection zone where the strong differential rotation discovered 
by helioseismology stretches out the poloidal
field to generate the toroidal field.  Then this toroidal field rises to the solar
surface due to magnetic buoyancy to produce the tilted bipolar sunspots.  The decay
of these tilted bipolar sunspots then gives rise to the poloidal field near the surface
by the Babcock--Leighton mechanism.  The meridional circulation is also indicated in
Figure~2.  We observe the meridional circulation to be poleward in the top layers of
the convection zone.  In order to conserve mass, the meridional circulation has to
be equatorward deeper down.  It is generally assumed in flux transport dynamo models
that the equatorward flow is at the bottom of the convection zone, although this is
not yet confirmed from observations. The poloidal field produced near the surface is
advected poleward by the poleward meridional circulation there, whereas the toroidal
field produced at the bottom of the convection zone is advected equatorward by the
equatorward meridional circulation there. This provides the theoretical explanation
of both the observed poleward drift of the surface magnetic field (outside active
regions) and the equatorward migration of the sunspots which form from the toroidal
field. While the basic idea of the flux transport dynamo was given in an early
paper by Wang, Sheeley and Nash [7], the first two-dimensional models were
constructed by Choudhuri, Sch\"ussler and Dikpati [8] and Durney [9].

A numerical code SURYA was developed in our group in Indian Institute of Science
to solve the basic equations of the flux transport dynamo [24--25] 
and was made public from 2005. A comparison
of the observational data shown in Figure~2 [10] with theoretical
results of [25] shown in Figure~10 makes it clear
that the flux transport dynamo is reasonably successful in reproducing various
aspects of the periodic behaviour of the sunspot cycle.  Apart from solving the
solar dynamo problem, the code SURYA has also been modified to study the accretion
of matter on magnetized neutron stars [26--27].
It may be noted that a flux tube approach has to be combined with the mean field
dynamo equation to have a more complete understanding of the magnetic field dynamics
within the solar convection zone [28]. For example, we have to consider
the wrapping of poloidal field lines around rising flux tubes to explain how
the observed current helicity of sunspots arise [29--30]. The flux transport dynamo model has also
been applied to model the back-reactions of the dynamo-generated magnetic field
such as torsional oscillations [31].

There have been some recent claims that the equatorward
reverse flow of the meridional circulation occurs at a shallow depth and not at
the bottom of the convection zone as usually assumed in the flux transport dynamo
model [32--33]. If these claims are corroborated by
other independent studies and turn out to be true, then we have to address the
question whether the flux transport dynamo can work with a shallow meridional
circulation.  Guerrero and de Gouveia Dal Pino [34] considered a shallow cell
of meridional circulation with equatorward turbulent pumping in the region below
it and succeeded in getting realistic butterfly diagrams.  Whether such latitudinal
pumping exists is questionable.  If there is just a shallow cell of meridional
circulation and nothing below it, then the flux transport dynamo cannot work.
However, recently Hazra, Karak and Choudhuri [35] showed that many of the attractive
features of the flux transport dynamo are retained if, below the shallow cell of
meridional circulation at the top of the convection zone, there are additional
cells such that there is an equatorward meridional circulation at the bottom of
the convection zone.  Thus, even if the meridional circulation has a return flow
at a shallow depth, the flux transport dynamo can presumably still work as long
as there is an appropriate equatorward flow at the bottom of the convection zone.

The original flux transport dynamo model of Choudhuri, Sch\"ussler
and Dikpati [8] led to two offsprings: a high diffusivity model
and a low diffusivity model.  The diffusion times in these two models
are of the order of 5 years and 200 years respectively.  The high
diffusivity model has been developed by a group working in IISc Bangalore
(Choudhuri, Nandy, Chatterjee, Jiang, Karak), whereas the low diffusivity
model has been developed by a group working in HAO Boulder (Dikpati,
Charbonneau, Gilman, de Toma).  The differences between these models
have been systematically studied by Jiang, Chatterjee and Choudhuri
[36] and Yeates, Nandy and Mckay [37].  Both these models are
capable of giving rise to oscillatory solutions resembling solar
cycles.  However, when we try to study the irregularities of the
cycles, the two models give completely different results.  We need
to introduce fluctuations to cause irregularities in the cycles.
In the high diffusivity model, fluctuations spread all over the
convection zone in about 5 years.  On the other hand, in the low
diffusivity model, fluctuations essentially remain frozen during
the cycle period.  Thus the behaviours of the two models are totally
different on introducing fluctuations.  As we shall see in the next
three Sections, only the high diffusivity model can provide explanations
for certain aspects of sunspot cycle irregularities. The high diffusivity
also helps in establishing the dipolar parity of the solar magnetic
field [25, 38]
and can explain the lack of significant hemispheric asymmetry 
[39--40].

\section{Nonlinearities versus random fluctuations}

The magnetic fields produced by the dynamo can react back on the 
velocity fields driving the dynamo action.  This leads to nonlinearities
in the mathematical theory.
It is well known that nonlinear dynamical systems can show complicated
chaotic behaviours and one possibility is that irregularities of the
sunspot cycle are just a manifestation of such chaotic behaviour. 
However, the mean field theory of the dynamo involves averaging
over turbulence and we always have fluctuations around the mean.
These random fluctuations also may be the source of irregularities.  For
some time, there has been a debate in this field whether the irregularities
of the sunspot cycle are primarily due to nonlinear chaos or due to
random fluctuations.  While we now think that there are signatures of
both the effects, it seems that the really large irregularities like
the grand minima are caused by random fluctuations.

Let us point out why many of us think that the largest irregularities
of the sunspot cycle are not due to nonlinear chaos.  The simplest
way of capturing the effect of the nonlinear feedback in a kinematic dynamo model (in
which the fluid equations are not solved) is to consider a quenching
of the $\alpha$ parameter (the crucial parameter in the dynamo
generation of magnetic fields) as follows:
$$\alpha = \frac{\alpha_0}{1 + |\overline{B}/B_0|^2}, \eqno(1)$$
where $\overline{B}$ is the mean magnetic field produced
by the dynamo and $B_0$ is the value of magnetic field beyond
which nonlinear effects become important.  There is a long history
of dynamo models studied with such quenching [41--43]. In most of the
nonlinear calculations, however, the dynamo eventually settles to a
periodic mode with a given amplitude rather than showing sustained
irregular behaviour.  The reason for this is intuitively obvious. 
Since a sudden increase in the amplitude of the magnetic field would
diminish the dynamo activity by reducing $\alpha$ given by (2) and thereby pull
down the amplitude again (a decrease in the amplitude would do the
opposite), the $\alpha$-quenching mechanism tends to lock the system
to a stable mode once the system relaxes to it. Only by using somewhat
unusual kinds of nonlinearities, usually with large time delays, it
is sometimes possible to get chaotic behaviour in the system.
Although nonlinearities may not produce sustained chaotic behaviour,
It has been suggested that the Gnevyshev--Ohl rule is caused by a period doubling
due to nonlinearities [44--45] and there is no other good
theoretical explanation for it. Presumably the nonlinearities play
some role in producing such effects as the Gnevyshev--Ohl rule, but we
believe that they are not the main cause behind the large irregularities
of the sunspot cycle.

Now let us come to the possibility that the irregularities of the
sunspot cycle are primarily caused by random fluctuations, as suggested
first by Choudhuri [46] and Hoyng [47]. The crucial issue is to figure
out the nature of random fluctuations in the flux transport dynamo. 
Choudhuri, Chatterjee and Jiang [48] identified the Babcock--Leighton mechanism of poloidal
field generation as the main source of random fluctuations.  This mechanism
depends on the tilts of bipolar sunspot pairs.  While the average tilts
are given by Joy's law, one finds a large scatter around this average, presumably
produced by the fact that the rising flux tubes are buffeted by turbulence in the \cz 
[49].  This scatter
around Joy's law produces fluctuations in the poloidal field generation process, 
ultimately giving rise to irregularities in the dynamo mechanism. In the high diffusivity
flux transport dynamo model, we can theoretically explain the observed correlation
between the polar field during the sunspot minimum and the strength of the
next cycle if the irregularities of cycles primarily arise due to fluctuations
in the Babcock-Leighton mechanism, but we do not get this correlation in the
low diffusivity model [36]. Since the origin
of this correlation in high diffusivity model has been discussed in detail in the
earlier review [10], we shall not get into a detailed discussion of this subject here,
except to mention that the theoretical explanation of this correlations lends support
simultaneously to the high diffusivity dynamo model and to the idea that the fluctuations in
the Babcock--Leighton mechanism is the major cause of irregularities in the sunspot
cycle.  Recent analyses of the sunspot tilt data by different groups also provide 
strong support to the scenario outlined above [50--51].

We have already mentioned that the correlation between the polar field
during a sunspot minimum and the strength of the next cycle provides a mechanism for
predicting future cycles.  We shall only make some comments on this. 
On the basis of the observation that the polar field was rather weak during the
last sunspot minimum, several groups predicted a few years ago
that the present cycle~24 would be rather weak [52--53]. One crucial 
question at that time was whether theoretical solar dynamo models
could be used to make a prediction.  During the sunspot minimum before the previous
cycle~23 (in the mid-1990s), solar dynamo models were still too primitive for
this purpose.  The sunspot minimum before the present cycle~24 was the first
sunspot minimum during which the solar dynamo models has reached a certain level of
sophistication to make such predictions.  Dikpati and Gilman [54] used their low
diffusivity model to predict that the cycle~24 would be the strongest cycle in 
the last half century.  On the other hand, Choudhuri, Chatterjee and Jiang [48]
used their high diffusivity model to predict that the cycle~24 will be the weakest
cycle in nearly a century.  This is a rather robust prediction of this high diffusivity model, because
this model produces a strong correlation between the polar field during the sunspot
minimum and the next cycle, and the fact that the polar field was very weak during
the last sunspot minimum was incorporated in the theoretical model for this prediction
work.  Figure~3 shows the present status of the sunspot number data with the two
theoretical predictions indicated. It is clear that the observational data is consistent
with the prediction of Choudhuri, Chatterjee and Jiang [48], making this the first
successful prediction of a cycle from a theoretical dynamo model in the history of
this subject. 

\begin{figure}
\center
\includegraphics[width=10cm]{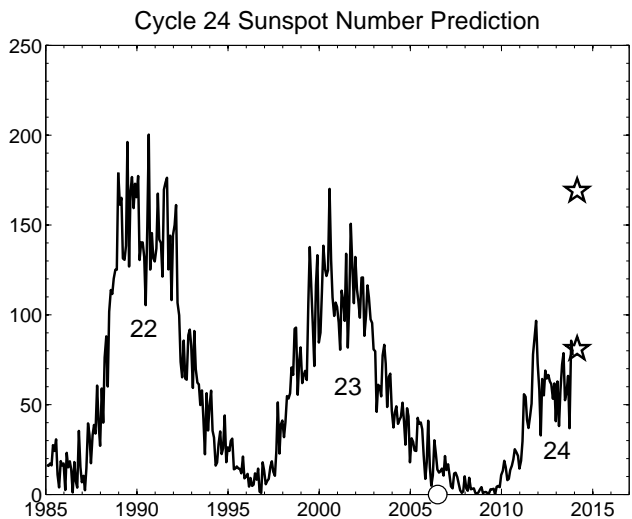}
  \caption{The monthly sunspot number plot for the last few years, indicating the
theoretical predictions. The upper star is the peak of cycle~24 predicted by
Dikpati and Gilman [54], whereas the lower star is what was predicted by Choudhuri,
Chatterjee and Jiang [48]. The circle on the horizontal axis indicates the time when
these predictions were made (in 2006).}
\end{figure}

Lastly, we come to the question whether fluctuations in the poloidal field generation
can produce grand minima.  When the poloidal field at the end of a cycle falls to
a very low value due to these fluctuations, Choudhuri and Karak [55] found that
the dynamo can be pushed into a grand minimum.  In fact, they were able to construct
an example of a grand minimum having the broad features of the Maunder minimum.
We thus conclude that the fluctuations in the Babcock--Leighton mechanism for
generating the poloidal field is a possible mechanism for producing grand 
minima---especially if the dynamo is not too supercritical [56].

\section{Fluctuations in meridional circulation}

Until about 5--6 years ago, it was not generally recognized that there is
another important source of sunspot cycle irregularities: fluctuations in
the \mc.  It is well known that the period of the flux transport dynamo varies roughly as
the inverse of the \mc\ speed.  The period of the dynamo is approximately given
by the time taken by \mc\ at the bottom of the convection zone to move from
higher latitudes to lower latitudes.  
Since the \mc\ determines the period of the flux transport dynamo, 
it is not surprising that any fluctuations
in \mc\ would have an effect on the flux transport dynamo. It has been found
recently that the \mc\ has a periodic variation with the solar cycle, becoming
weaker at the time of sunspot maximum [57--59].
Presumably the Lorentz force of the dynamo-generated magnetic field slows down
the \mc\ at the time of the sunspot maximum.  Karak and Choudhuri [60] found
that this quenching of \mc\ by the Lorentz force does not produce irregularities
in the cycle, provided the diffusivity is high as we believe.  Then the question
arises whether there are other kinds of fluctuations in the \mc\ apart from
these cyclic modulations.

We have reliable observational data on the variation of \mc\ only for a little
more than a decade.  To draw any conclusions about the variation of \mc\ at
earlier times, we have to rely on indirect arguments. If we assume the cycle
period to go inversely as \mc, then we can use periods of different past solar
cycles to infer how \mc\ has varied with time in the last few centuries. On the
basis of such considerations, it appears that the \mc\ had random fluctuations
in the last few centuries with correlation time of the order of 30--40 years [61].
We now come to question what effect these random fluctuations of \mc\ may have
on the dynamo.  Based on the analysis of Yeates, Nandy and Mckay [37], we can
easily see that dynamos with high and low diffusivity will be affected very
differently.  Suppose the \mc\ has suddenly fallen to a low value. This will
increase the period of the dynamo and lead to two opposing effects.  On the one
hand, the differential rotation will have more time to generate the toroidal field and
will try to make the cycles stronger.  On the other hand, diffusion will also have
more time to act on the magnetic fields and will try to make the cycles weaker.
Which of these two competing effects wins over will depend on the value of diffusivity.
If the diffusivity is high, then the action of diffusivity is more important and
the cycles become weaker when the \mc\ is slower.  The opposite happens if the
diffusivity is low.

The important question now is if there is any kind of observational data to indicate
whether the cycles become weaker (which will happen for high diffusivity) or stronger
(which will happen for low diffusivity) when the \mc\ is slower.  
The Waldmeier effect discussed in \S2 provides precisely this kind of observational
data. The rise time of the sunspot cycle roughly goes as the duration of the cycle.  If the
\mc\ is slower, then the cycle is longer and the rise time is also longer.  According
to the Waldmeier effect, the longer cycle tends to be weaker in strength.  This 
happens only if the turbulent diffusivity is high.  Karak and Choudhuri [61] were
able to explain the Waldmeier effect on the basis of the high diffusivity model,
whereas the low diffusivity would give the opposite of the Waldmeier effect.  The
success in explaining the Waldmeier effect is another feather in the cap of the
high diffusivity model.

Since a slowing
of the \mc\ would make the cycles weaker, a question that comes before us is
whether a sufficient slowing of the \mc\ can cause a grand
minimum.  Karak [62] indeed found that the flux transport
dynamo can be pushed into a grand minimum if the \mc\ drops to 0.4  of its normal
value.  This is clearly another possible mechanism for producing a grand minimum.

\section{A theoretical model of grand minima}

From the discussions in the previous two sections, it should be clear that a grand
minimum can be caused by two means: if the poloidal field produced at the end of cycle 
is very weak as a result of fluctuations in the Babcock--Leighton mechanism and if 
the \mc\ falls to a very low value due to its fluctuations. Presumably the grand
minima arise due to the combined effect of both these kinds of fluctuations, as shown
by Choudhuri and Karak [63]. Let $\gamma$ be the normalized strength of the polar
field (i.e.\ the strength of the polar field divided by its average value over many
cycles) at the end of a cycle and let $v_0$ be the amplitude of the \mc. Figure~4 shows
the two-dimensional parameter space of $\gamma$ versus $v_0$.  The condition at the
beginning of a sunspot cycle is clearly represented by a point in this two-dimensional
parameter space. Choudhuri and Karak [63] found that the dynamo is pushed into a
grand minimum if the condition at the beginning of the cycle corresponds to the shaded 
region of the parameter space.  What is the probability of this happening?

\begin{figure}
\center
\includegraphics[width=9cm]{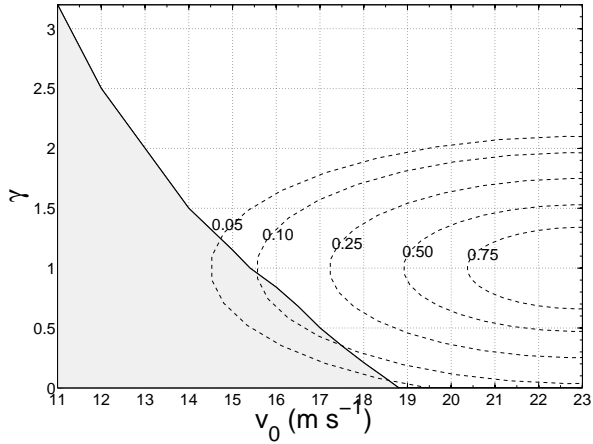}
\caption{The parameter space indicating the normalized strength $\gamma$ of the
poloidal field and the amplitude of the \mc, the shaded region being the
part to the parameter space giving rise to
grand minima.}
\end{figure}

Presumably both the fluctuations we are considering would be of Gaussian nature.
Then the joint probability that the polar field strength at the
end of a cycle lies in the range $\gamma$, $\gamma + d\gamma$ and the
amplitude of the \mc\ at the same time lies in the range $v_0$, $v_0 +dv_0$ is given by
$$P (\gamma, v_0) d\gamma dv_0 = \frac{1}{\sigma_v \sqrt{2 \pi}} \exp\left[- \frac{(v_0 - \overline{v_0})^2}{2 \sigma_v^2}\right]
\frac{1}{\sigma_{\gamma} \sqrt{2 \pi}} \exp\left[- \frac{(\gamma - 1)^2}{2 \sigma_{\gamma}^2}\right] d\gamma \, dv_0. \eqno(2)$$
The probability that the condition at the beginning of a cycle lies in the shaded region
of Figure~4 is obtained by integrating the double Gaussian given by (2) over this region.
To carry on this integration, we need values of $\sigma_v$ and $\sigma_{\gamma}$, which are
the widths of these Gaussians.  Choudhuri and Karak [63] realized that these can be obtained
from the observational data of the last 28 cycles.  The periods of these cycles give the
values of the \mc\ during these cycle, from which the probability distribution function of the
\mc\ can be constructed.  Since strengths of the cycles are correlated with the polar field
strength $\gamma$ at the beginning of the cycle, the strengths of the last 28 cycles can
be used to construct the probability distribution function of $\gamma$.  Although we would
not expect a very good Gaussian fit from a set of 28 data points, Choudhuri and Karak [63]
found that the fits were not too bad and could estimate the values of $\sigma_v$, $\sigma_{\gamma}$.
On carrying out the integration of the double Gaussian over the shaded region in Figure~4,
Choudhuri and Karak [63] found the value to be 1.7\%.  This means that 17 cycles out of
1000 cycles (in 11,000 years) would have conditions appropriate for grand minima at their
beginnings.  This is remarkably close to the observational data that there had been 27 grand
minima in the last 11,000 years. In fact, in actual runs of the dynamo code with fluctuations
given by the double Gaussian (2), Choudhuri and Karak [63] typically found about 24--30 grand
minima in a run spanning 11,000 years. 

While this may seem like a very encouraging result, one aspect of grand minima still remains
completely shrouded in mystery.  If there are no sunspots at all during a grand minimum,
one important question is whether the Babcock--Leighton mechanism which depends on the
existence of tilted bipolar sunspots can operate at all.  If the Babcock--Leighton mechanism
is not operative, then some mechanism has to build up the poloidal field so that the Sun
can eventually come out of the grand minimum.  If the magnetic field during the grand
minimum becomes sufficiently weak, then one possibility is that the $\alpha$-effect originally
envisaged by Parker [3] and Steenbeck, Krause and R\"adler [4] becomes operative.
Karak and Choudhuri [64] have done some explorations of this.  The results are inconclusive.
While we now have some idea how the Sun gets pushed into grand minima, we have very little
understanding how the Sun gets out of a grand minimum after falling into one. 

\section*{Acknowledgement}

I thank DST for partial support through a J C Bose Fellowship.


\def\apj{{\it Astrophys.\ J.}}
\def\mnras{{\it Mon.\ Notic.\ Roy.\ Astron.\ Soc.}}
\def\sol{{\it Solar Phys.}}
\def\aa{{\it Astron.\ Astrophys.}}
\def\gafd{{\it Geophys.\ Astrophys.\ Fluid Dyn.}}


\begin{thebibliography}{99}

\bibitem[1]{sch44}
  S H Schwabe,   {\it Astron.\ Nachr.} {\bf 21}, 233 (1844)

\bibitem[2]{hale08}
  G E Hale,   \apj\ {\bf 28}, 315 (1908)

\bibitem[3]{par55b}
  E N Parker,  \apj\ {\bf 122}, 293 (1955)
 
\bibitem[4]{ste66}
  M Steenbeck, F Krause and K-H R\"adler,  {\it Z.\ 
  Naturforsch.} {\bf 21a}, 1285 (1966)

\bibitem[5]{bab61}
  H W Babcock,  \apj\ {\bf 133}, 572 (1961)

\bibitem[6]{lei69}
  R B Leighton,  \apj\ {\bf 156}, 1 (1969)

\bibitem[7]{wan91}
  Y-M Wang, N R Sheeley and A G Nash,  \apj\ {\bf 383}, 431 (1991)

\bibitem[8]{cho95}
  A R Choudhuri, M Sch\"ussler and M Dikpati,  \aa\
  {\bf 303}, L29 (1995)

\bibitem[9]{dur95}
  B R Durney,  \sol\ {\bf 160}, 213 (1995)

\bibitem[10]{chou11}
  A R Choudhuri,  {\em Pramana} {\bf 77}, 77 (2011)

\bibitem[11]{gne48}
  M N Gnevyshev and A I Ohl, {\em Astron.\ Zh.} {\bf 25}, 18 (1948)

\bibitem[12]{uso07}
  I G Usoskin, S K Solanki and G A Kovaltsov, \aa\ {\bf 471}, 301 (2007)

\bibitem[13]{wald35}
  M Waldmeier, {\em Mitt.\ Eidgen.\ Sternw.\ Zurich} {\bf 14}, 105 (1935)

\bibitem[14]{spr81}
  H C Spruit,  \aa\ {\bf 98}, 155 (1981) 

\bibitem[15]{chou90}
  A R Choudhuri,  \aa\ {\bf 239}, 335 (1990)

\bibitem[16]{cho87}
  A R Choudhuri and P A Gilman,  \apj\ {\bf 316}, 788 (1987)

\bibitem[17]{cho89}
  A R Choudhuri,  \sol\ {\bf 123}, 217 (1989)

\bibitem[18]{sil93}
  S D'Silva and A R Choudhuri,  \aa\ {\bf 272}, 621 (1993)

\bibitem[19]{fan93}
  Y Fan, G H Fisher and E E DeLuca,  \apj\ {\bf 405}, 390 (1993)

\bibitem[20]{wan89}
  Y-M Wang, A G Nash and N R Sheeley,  \apj\
  {\bf 347}, 529 (1989)

\bibitem[21]{dik94}
  M Dikpati and A R Choudhuri, \aa\ {\bf 291}, 975 (1994)

\bibitem[22]{dik95}
  M Dikpati and A R Choudhuri, \sol\ {\bf 161}, 9 (1995)

\bibitem[23]{cho99}
  A R Choudhuri and M Dikpati, \sol\ {\bf 184}, 61 (1999)

\bibitem[24]{nan02}
  D Nandy and A R Choudhuri,  {\it Science} {\bf 296}, 1671 (2002)

\bibitem[25]{cha04}
  P Chatterjee, D Nandy and A R Choudhuri,  \aa\ {\bf 427}, 1019 (2004)

\bibitem[26]{cho02}
  A R Choudhuri and S Konar, \mnras\ {\bf 332}, 933 (2002)

\bibitem[27]{cho04}
  S Konar and A R Choudhuri, \mnras\ {\bf 348}, 661 (2004)

\bibitem[28]{cho03}
  A R Choudhuri, \sol\ {\bf 215}, 31 (2003)

\bibitem[29]{cho04}
  A R Choudhuri, P Chatterjee and D Nandy,  \apj\ {\bf 615}, L57 (2004)

\bibitem[30]{cha04}
  P Chatterjee, A R Choudhuri and K Petrovay,  \aa\ {\bf 449}, 781 (2006)

\bibitem[31]{cha09}
  S Chakraborty, A R Choudhuri and P Chatterjee, {\em Phys.\ Rev.\ Lett.} {\bf 102}, 041102 (2007)

\bibitem[32]{hatha12}
  D H Hathaway, \apj\ {\bf 760}, 84 (2012)

\bibitem[33]{zhao13}
  J Zhao, R S Bogart, A G Kosovichev, T L Duvall and T Hartlep, \apj\ {\bf 774}, L29 (2013)

\bibitem[34]{Guerr08}
  G Guerrero and E M de Gouveia Dal Pino, \aa\ {\bf 485}, 267 (2008)

\bibitem[35]{haz14}
  G Hazra, B B Karak and A R Choudhuri, \apj\, submitted, arXiv:1309.2838 (2014)

\bibitem[36]{jia07}
  J Jiang, P Chatterjee and A R Choudhuri, \mnras\ {\bf 381}, 1527 (2007)

\bibitem[37]{yea08}
  A R Yeates, D Nandy and D H Mackay, \apj\  
  {\bf 673}, 544 (2008)
  
\bibitem[38]{hot10b}
  H Hotta and T Yokoyama, \apj\ {\bf 714}, L308 (2010)

\bibitem[39]{cha06} 
  P Chatterjee and A R Choudhuri, \sol\ {\bf 239}, 29 (2006)

\bibitem[40]{goe09} 
  A Goel and A R Choudhuri,  {\em Res.\ Asron.\ Astrophys.} {\bf 9}, 115 (2009)

\bibitem[41]{stix72}
  M Stix, \aa\ {\bf 20}, 9 (1972)

\bibitem[42]{ivan77}
  T S Ivanova and A A Ruzmaikin, {\it Soviet Astron.} {\bf 21}, 479 (1977)

\bibitem[43]{yosh78}
  H Yoshimura, \apj\ {\bf 226}, 706 (1978)

\bibitem[44]{cha05}
  P Charbonneau, C St-Jean and P Zacharias, \apj\ {\bf 619}, 613 (2005)

\bibitem[45]{cha06}
  P Charbonneau, G Beaubien and C St-Jean, \apj\ {\bf 658}, 657 (2007)

\bibitem[46]{cho92}
  A R Choudhuri,  \aa\ {\bf 253}, 277 (1992)

\bibitem[47]{hoy93}
  P Hoyng, \aa\ {\bf 272}, 321 (1993)

\bibitem[48]{cho07}
  A R Choudhuri, P Chatterjee and J Jiang, {\em Phys.\ Rev.\ Lett.} 
  {\bf 98}, 131103 (2007)

\bibitem[49]{lon02}
  D Longcope and A R Choudhuri,  \sol\ {\bf 205}, 63 (2002)

\bibitem[50]{dasi10}
  M Dasi-Espuig, S K Solanki, N A Krivova, R Cameron and T Pe\~nuela,  \aa\ {\bf 518}, 7 (2010)

\bibitem[51]{kitch11}
  L L Kitchatinov and S V Olemskoy, {\em Astronomy Letters} {\bf 37}, 656 (2011)

\bibitem[52]{sva05} 
  L Svalgaard, E W Cliver and Y Kamide, 
  \textit{Geo. Res. Lett.} {\bf 32}, L01104 (2005)

\bibitem[53]{sch05} 
  K Schatten, \textit{Geo. Res. Lett.} {\bf 32}, L21106 (2005)

\bibitem[54]{dik06}
  M Dikpati and P A Gilman,  \apj\ {\bf 649}, 498 (2006)

\bibitem[55]{cho09}
  A R Choudhuri and B B Karak, {\em Res.\ Asron.\ Astrophys.} {\bf 9}, 953 (2009)

\bibitem[56]{olem13}
  S V Olemskoy, A R Choudhuri and L L Kitchatinov, {\em Astronomy Reports} {\bf 57}, 458 (2013)

\bibitem[57]{chou01}
  D-Y Chou and D-C Dai, \apj\	{\bf 559}, L175 (2001)

\bibitem[58]{hr10}
  D H Hathaway and L Rightmire, 2010, {\em Science} {\bf 327}, 1350 (2010)

\bibitem[59]{ba10}
  S Basu and H M Antia, 2010, \apj\ {\bf 717}, 488 (2010)

\bibitem[60]{kar12}
  B B Karak and A R Choudhuri, \sol\ {\bf 278}, 137 (2012)

\bibitem[61]{kar11}
  B B Karak and A R Choudhuri, \mnras\ {\bf 410}, 1503 (2011)

\bibitem[62]{kar10}
  B B Karak, \apj\ {\bf 724}, 1021 (2010)

\bibitem[63]{cho12}
  A R Choudhuri and B B Karak, {\em Phys.\ Rev.\ Lett.} {\bf 109}, 171103 (2012)

\bibitem[64]{kar13}
  B B Karak and A R Choudhuri, {\em Res.\ Asron.\ Astrophys.} {\bf 13}, 1339 (2013)

 

\end{thebibliography}
\end{document}